\documentclass[superscriptaddress,
bibnotes,amsmath,amssymb,aps,pra,floatfix,twocolumn]{revtex4-2}

\pdfoutput=1

\usepackage{amssymb}
\usepackage{amsmath}
\usepackage{amsthm}
\usepackage{graphicx,bm}
\usepackage{epstopdf}
\usepackage{subfigure}
\usepackage{natbib}
\usepackage{epsfig}
\usepackage{amsfonts}
\usepackage{mathrsfs}
\usepackage{CJK}
\usepackage{array,longtable}
\usepackage{float}
\usepackage[usenames, dvipsnames, x11names]{xcolor}
\usepackage[colorlinks]{hyperref}
\usepackage[capitalise]{cleveref}
\usepackage{enumerate}
\usepackage{soul}
\usepackage{dsfont}
\hypersetup{
 colorlinks=true,
 citecolor=red,
 linkcolor=blue,
 urlcolor=cyan}
\usepackage{mathtools}
\usepackage{lipsum}

\newcommand{\tr}{\mathrm{tr}}
\newcommand{\sgn}{ \ \mathrm{sgn}}

\renewcommand{\L}{\hat{\mathcal{L}}}

\newcommand{\D}{\mathcal{D}}

\newcommand{\1}{\mathds{1}}

\newcommand{\hc}{\mathrm{H.c.}}
\newcommand{\rss}{\hat{\rho}_\mathrm{ss}}

\newcommand{\Erel}{\mathcal{E}_{2}}
\newcommand{\var}{\mathrm{var}}

\usepackage{comment}

\begin{document}

\title{Accelerating Dissipative State Preparation with Adaptive Open Quantum Dynamics}
\author{Andrew Pocklington}
\email{abpocklington@uchicago.edu}
\affiliation{Department of Physics, University of Chicago, 5640 South Ellis Avenue, Chicago, Illinois 60637, USA }

\affiliation{Pritzker School of Molecular Engineering, University of Chicago, Chicago, IL 60637, USA}

\author{Aashish A. Clerk}
\email{aaclerk@uchicago.edu}
\affiliation{Pritzker School of Molecular Engineering, University of Chicago, Chicago, IL 60637, USA}

\date{\today}

\begin{abstract}
A wide variety of dissipative state preparation schemes suffer from a basic time-entanglement tradeoff: the more entangled the steady state, the slower the relaxation to the steady state.  Here, we show how a minimal kind of adaptive dynamics can be used to completely circumvent this tradeoff, and allow the dissipative stabilization of maximally entangled states with a finite time-scale.  Our approach takes inspiration from simple fermionic stabilization schemes, which surprisingly are immune to entanglement-induced slowdown.  We describe schemes for accelerated stabilization of many-body entangled qubit states (including spin squeezed states), both in the form of discretized Floquet circuits, as well as continuous time dissipative dynamics.  Our ideas are compatible with a number of experimental platforms.    
\end{abstract}


\maketitle 


\textit{Introduction---}
There is renewed interest in using the power of non-unitary time evolution to prepare complex entangled states of quantum many-body systems.  Protocols based on measurements and classical feedback \cite{Ahn2002,Riste2013}, as well as on using engineered dissipation 
\cite{Poyatos1996, Plenio2002}, have received considerable theoretical and experimental attention.  Of particular interest are stabilization schemes whose dynamics have unique, highly-entangled steady states.  Such schemes have numerous advantages over more conventional approaches, and are particularly intriguing when the goal is establishing large amounts of remote entanglement \cite{Stannigel2012,ZipilliPRL2013,Doucet2020,Govia2022,Brown2022,Angeletti2023,lingenfelter2023,Diehl2008,Kraus2008,Diehl2011,Pocklington2022,Zippilli2015,Ma2017}.  

For all non-unitary stabilization protocols, it is crucial both to have a useful steady state, as well as a relaxation rate that is fast.  If relaxation is too slow, then other unwanted dissipative dynamics will invariably degrade the final stabilized state.  For dissipative remote entanglement stabilization, a variety of different schemes all suffer from a basic time-entanglement tradeoff: the more entangled the steady state, the slower the relaxation \cite{Doucet2020,Govia2022,Pocklington2022,Brown2022,lingenfelter2023,Agusti2023}.  Recent work \cite{Pocklington2024} unified these results, proving a fundamental time-entanglement bound that constrains a wide set of local dissipative stabilization protocols.
In particular, the stabilization timescale diverges for maximally entangled states.
\begin{figure}[t]
    \centering
    \includegraphics[width=\linewidth]{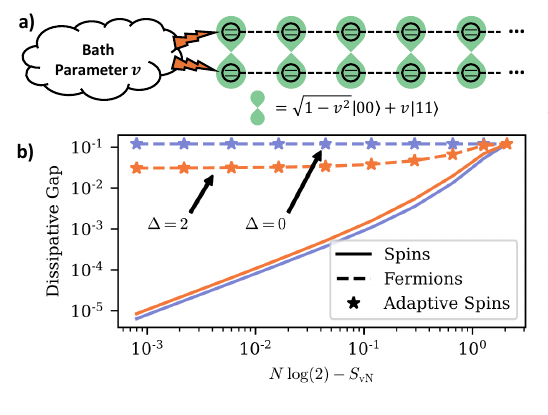}
    \caption{
    a) Schematic showing two chains of $n$ qubits/ fermions each that only interact via coupling to a common engineered bath at the boundary.  Tuning the bath parameter $v$ tunes the steady state interchain entanglement $S_\mathrm{vN} \leq N \log 2$. b) Dissipative gap vs $S_\mathrm{vN}$ for $2n=6$ total fermions (dashed lines), spins (solid lines) and adaptive spins (stars). 
    Qubits exhibit 
    a dramatic slowdown with increasing entanglement, fermions do not.  Dynamics is given by \cref{eqn:XXf,eqn:jump_f} (fermions), \cref{eqn:XXZs,eqn:local_jump} (spins), with $J = \Gamma = 1$, and interaction $\Delta = 0$ (blue) or $\Delta = 2$ (orange). The adaptive spin protocol (given by \cref{eqn:XXZs,eqn:time_cts_jumps}) mimics the fermionic model, with no entanglement induced slowdown.   
    }
    \label{fig:4}
\end{figure}

These studies raise a natural question: can we use additional resources to speed up entanglement stabilization? One method is to use only a subset of the total available Hilbert space, so as to never try to reach the maximal amount of entanglement \cite{Brown2022}. If instead one wants to use only spin degrees of freedom and fully prepare the maximal number of Bell pairs, we show that this is possible in many settings by using a minimal kind of adaptive dynamics. 

 Adaptive dynamics whereby one does measurement and feedback on a quantum system has been explored both theoretically (e.g., \cite{Wiseman2009,Ivanov2020,Nakagawa2024,Piroli2021,Tantivasadakarn2022,Lu2022}) as well as experimentally (e.g., \cite{Kroeger2020,Gajdacz2016,Leroux2012,Sayrin2011,Vijay2012,Magrini2021}) where it as been successful in preparing quantum states or phases.

We start with the realization that simple fermionic entanglement stabilization schemes are not limited by the time-entanglement tradeoff that plagues most qubit schemes.  Formally, the fact that spatially distinct fermionic operators need not commute means that the results of Ref.~\cite{Pocklington2024} do not apply.  More physically, fermionic statistics provide a kind of built-in non-locality that is, surprisingly, a resource for entanglement stabilization.  We show that the basic ingredients that give fermions an advantage can be mimicked in qubit systems with local interactions, augmented by a minimal classical memory that time-correlates subsequent quantum jumps.  This dramatically accelerates the preparation of maximally entangled states (as well as spin squeezed states), both in time-continuous schemes, as well as in discrete circuits (see Fig.~\ref{fig:4}).

\textit{Fermions evade slowdown---}
We start by recapping the result of \cite{Pocklington2024}. Consider a system defined on a bipartite Hilbert space $\mathcal{H} = \mathcal{H}_A \otimes \mathcal{H}_B$ with local Hilbert space dimension $\dim \mathcal{H}_A = \dim \mathcal{H}_B = N$. 
We assume that the subsystems are physically separated, but interact with a common Markovian reservoir that mediates intersystem interactions (c.f.~\cref{fig:4}). The system density matrix is then described by a Lindblad
master equation $\partial_t \hat \rho = \L \hat \rho$ whose form satisfies
\begin{align}
    \L &= -i[\hat H, \bullet] + \sum_\mu \D[\hat L_\mu], \label{eqn:master_equation} \\
    \hat L_\mu &= \hat A_\mu \otimes \1 + \1 \otimes \hat B_\mu.\label{eqn:locality_constraint}
\end{align}
Here, $\mathcal{D}[\hat L] \hat \rho = \hat L \hat \rho \hat L^\dagger - \frac{1}{2} \{ \hat L^\dagger \hat L, \hat \rho \}$. The sum-of-locals form in \cref{eqn:locality_constraint} reflects the locality constraint on the system-bath coupling  
\cite{Pocklington2024}. Next, suppose $\L$ has a pure steady state $\hat \rho_{\mathrm{ss}} = |\psi \rangle \langle \psi |$, i.e.~$\L \hat \rho_{\mathrm{ss}} = 0$. An initial state $\hat \rho_0$ evolves as $\hat \rho_t = e^{\L t} \hat \rho_0$ and its fidelity to the steady state is $F(t) = \left( \tr \sqrt{\sqrt{\rss} \hat \rho_t \sqrt{\rss}} \right)^2$. 
Defining $S^{(2)}_{\mathrm{ss}}$ to be the steady state Renyi-2 entanglement entropy between subsystems $A$ and $B$, Ref.~\cite{Pocklington2024} shows that 
$|\partial_t F(t)|$ is bounded above by $\Gamma_\mathrm{max} \propto \delta \Erel$ where $\delta \Erel = \left( e^{-S^{(2)}_{\mathrm{ss}} } - \frac{1}{N} \right)$ is a measure of how close the state is to being maximally entangled.  This implies that the  relaxation rate vanishes as the steady state entanglement approaches its maximal value.  

Given this general result for spins, one might assume it also applies to fermionic systems. There is, however, a loophole:  as fermions satisfy canonical \textit{anti}-commutation relations, operators localized to the $A$ system need not commute with ones localized on the $B$ system. Formally, the theorem of  Ref.~\cite{Pocklington2024} assumes a tensor product Hilbert space.  This is violated by fermions, as they live in the exterior algebra $\mathcal{H} = \mathcal{H}_A \wedge \mathcal{H}_B$ (the totally antisymmetrized part of the tensor product space \cite{Vourdas2018}). This structure implies a subtle kind of non-locality, something that becomes more explicit if we map a given fermionic dynamics to qubits. For example, the Jordan-Wigner transformation maps qubit operators $\hat \sigma^+, \hat \sigma^-$ to fermionic creation and annihilation operators $\hat c, \hat c^\dagger$ via
\begin{align}
    \hat c_i &= \left( \prod_{j = 1}^{i -1} \hat \sigma_{j}^z \right) \hat \sigma_i^-.
\end{align}
The Pauli string $\left( \prod_{j = 1}^{i -1} \hat \sigma_{j}^z \right)$ captures the non-local essence of the canonical fermionic anticommutation relations.  Hence, what appeared to be a local fermionic dissipative interaction (i.e.~satisfying \cref{eqn:locality_constraint}) will in general look highly non-local when mapped to qubits.  

\begin{figure*}[t]
    \centering
    \includegraphics{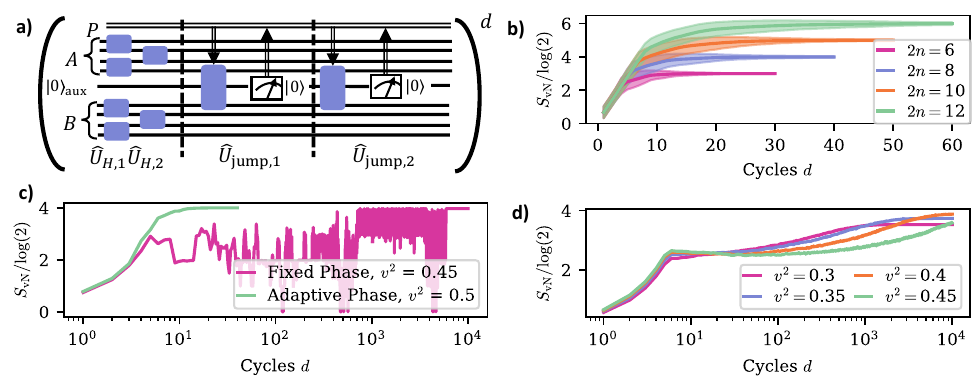}
    \caption{a) Brickwork adaptive circuit that allows for rapid entanglement stabilization by mimicking fermionic dynamics of  \cref{eqn:fermi_spin_1,eqn:fermi_spin_2}. Solid black lines: qubits, double black lines: classical single-bit memory,  blue rectangles:  unitary gates [c.f. \cref{eqn:circuitGatesHam,eqn:circuitGatesJump}].  Each cycle is repeated $d$ times. b) Average entanglement dynamics for the adaptive circuit with $v^2 = 0.5$, generating $n$ perfect bell pairs. $S_\mathrm{vN}$ is the von Neumann entanglement entropy between the $A$ and $B$ chains measured at the end of the cycle; solid lines shows average values over 1000 runs, and shaded area shows one standard deviation. c) Entanglement dynamics over a single trajectory for the adaptive circuit (green), as well as the non-adaptive variant where $P$ is fixed over the full run (pink). d) Average entanglement dynamics for a non-adaptive circuit with $2n=8$ and variable $v^2$. Each line is averaged over 2500 runs. In both b) and d) the system begins in the product state $|\psi \rangle = \otimes_{j=1}^{2n + 1} |0 \rangle^{}$ with $P=1$. For all plots,  $J=1$ in $\hat U_{H,1}$ and $\hat U_{H,2}$.}
    \label{fig:8}
\end{figure*}

We now show that this effective non-locality can be used as a resource to speed up fermionic entanglement generation. Consider two fermionic tight-binding chains with hopping $J$, locally coupled to a Markovian bath at their left boundaries (coupling rate $\Gamma$) 
(see \cref{fig:4}).  The bath is constructed so that it creates pairing correlations between the boundary sites (for other examples of linear-in-fermion jump operators, see e.g. \cite{Diehl2011,Shtanko2021}). The Lindbladian is $\L = -i[\hat H_{\mathrm{f}}, \bullet] + \L_{\mathrm{f}}$, with
\begin{align}
    \hat H_{\mathrm{f}} &= \sum_{s = A,B} \sgn(s) \left[ \sum_{i =1}^{n-1} J\hat c_{s,i}^\dagger \hat c_{s,i +1} + \frac{\Delta}{2} \hat n_{s,i} \hat n_{s,i + 1}  \right]+ \hc, \label{eqn:XXf} \\
    \L_{\mathrm{f}} &= \Gamma \D[u \hat c_{A,1} +  v \hat c_{B,1}^\dagger] + \Gamma \D[u \hat c_{B,1} - v \hat c_{A,1}^\dagger]. \label{eqn:jump_f}
\end{align}
Here, $\hat{n}_{\alpha} = \hat{c}^\dagger_{\alpha} \hat{c}_{\alpha}$, $u^2 + v^2 = 1$, and $\sgn(A) = -\sgn(B) = 1$. This master equation explicitly obeys the locality constraint of \cref{eqn:locality_constraint} (i.e.~jump operators are sums-of-locals).  There is a unique, pure steady state given by $|\psi \rangle = \prod_{i = 1}^n  \left(  u - v \hat c_{A,i}^\dagger \hat c_{B,i}^\dagger \right) |0\rangle$ where $|0 \rangle$ is the joint fermionic vacuum \cite{Pocklington2022}. As $u \to v$, the steady state of this system becomes maximally entangled, but the dissipative gap is independent of $u,v$ and remains $\mathcal{O}(\Gamma)$.  The relaxation timescale is thus independent of the amount of steady state entanglement, and this fermionic system has no entanglement-time tradeoff,  see \cref{fig:4}. Even for $\Delta \neq 0$ (i.e.~$\hat{H}_{\rm f}$ is now interacting), there is no slow down.  The steady state is unchanged, and the dissipative gap does not close with increasing steady-state entanglement, see \cref{fig:4}.

To emphasize that the fermionic nature of the above system was crucial, consider the behaviour of the analogous qubit setup, which is a pair of anisotropic XXZ spin chains with boundary pairing dissipation:
\begin{align}
    \hat H_\mathrm{s} &=   \sum_{s=A,B} \sgn(s) \left[ \sum_{i =1}^{n-1} J\hat \sigma_{s,i}^+ \hat \sigma_{s,i +1}^- + \frac{\Delta}{2} \hat \sigma_{s,i}^z \hat \sigma_{s,i + 1}^z \right] + \hc, \label{eqn:XXZs} \\
    \L_{\mathrm{s}} &=  \D[u \hat \sigma_{A,1}^- + v \hat \sigma_{B,1}^+] + \D[u \hat \sigma_{B,1}^- + v \hat \sigma_{A,1}^+]. \label{eqn:local_jump}
\end{align}
The master equation governed by $\L = -i[\hat H_{\mathrm{s}}, \bullet] + \L_{\mathrm{s}}$ also has a pure, entangled steady state given by $n$ dimerized pairs of spins: $|\psi \rangle = \prod_{i=1}^n \left(  u - v \hat  \sigma_{A,i}^+ \hat \sigma_{B,i}^+ \right) |0 \rangle$ in exact analogue to the fermions. However, the dissipative gap closes as predicted by the bound \cite{Pocklington2024}, see \cref{fig:4}. We are thus left with the conclusion that the effective non-locality inherent in fermionic dissipative evolution can be a resource for entanglement stabilization: they can completely evade the slowdown that is fundamental to analogous qubit setups.

\textit{Fermion-inspired adaptive dynamics---}
Surprisingly, we now show that knowing about the exceptional behaviour of fermionic dissipative models can be used to modify qubit models in a way that preserves locality, but dramatically speeds up entanglement stabilization.  Note first that adding dissipators with a jump operator of the form $\hat L = \hat A \otimes \hat B$ can circumvent the bound of Ref.~\cite{Pocklington2024}, as this no longer satisfies the locality condition \cref{eqn:locality_constraint}.  However, it is not {\it a priori} obvious what sort of dissipators should be added for speeding up the dynamics while preserving the steady state entanglement.    

Our study of fermions, however, provides a recipe for what to do: mimic the fermionic dynamics.  The most naive way to do this would be to directly implement the needed string operators in order to realize \cref{eqn:XXf,eqn:jump_f} with qubits:  
\begin{align}
    \hat H &= J \sum_{s = A,B} \sgn(s) \sum_{i = 1}^{n-1}  \hat \sigma_{s,i}^+ \hat \sigma_{s,i + 1}^- + \hc \label{eqn:fermi_spin_1}\\
    \hat L_1 &= u \hat \sigma_{A,1}^- - v \hat \sigma_{B,1}^+ (-1)^{\hat n}, \ \ \hat L_2 = u \hat \sigma_{B,1}^- (-1)^{\hat n} - v \hat  \sigma_{A,1}^+, \label{eqn:fermi_spin_2}
\end{align}
with $\L = -i[\hat H, \bullet] + \D[\hat L_1] + \D[\hat L_2]$.  Here, $\hat n$ is the total $Z$ magnetization in both chains $\hat n = \frac{1}{2}\sum_{s,i} ( \hat \sigma_{s,i}^z + 1)$. This qubit model will evade slow-down, as it is equivalent to the fermion master equation \cref{eqn:XXf,eqn:jump_f} under a Jordan-Wigner transformation (with $\Delta = 0$ for simplicity), \cite{Pocklington2022}. However, because of the string operators, it corresponds to a long-range dissipative interaction between the chains that would be extremely difficult to engineer. We will thus consider an alternate route.  

Note that in \cref{eqn:fermi_spin_2}, the parity operator $\hat P \equiv (-1)^{\hat n}$ is the only object that violates our locality constraint.  Further note that it yields a weak symmetry of the dynamics: $[\hat H, \hat P] = \{ \hat L_i, \hat P \} = 0$, \cite{Buca2012}.  More explicitly, the Hamiltonian conserves parity, and each quantum jump simply flips it sign.  This suggests a strategy: by starting in a definite parity state and then tracking the number of jumps that have occurred, we can treat $P = \langle \hat P \rangle$ as a 
{\it classical variable}.  We can use this variable to update the dynamics, by setting a phase in the system-bath coupling to match its value.  In doing so, we eliminate the need to engineer an explicit long-range interaction between the chains.  Instead, we have an example of an adaptive protocol. We stress that this ``fermion-inspired'' protocol is able to exponentially improve the preparation time of the steady state, while requiring the absolute minimal resource of a single bit classical memory necessary to evade the bounds set in \cite{Pocklington2024}.

While such an adaptive protocol may seem non-trivial to engineer in a time-continuous open quantum system, it can be done very efficiently via a measured unitary circuit. Adaptive quantum circuits have been shown to be extremely powerful in other contexts, including entanglement generation \cite{Piroli2021,Tantivasadakarn2022,Lu2022} and topological ground state preparation \cite{Verresen2022,Bravyi2022,Tantivasadakarn2023}. While the power of adaptive circuits is often tied to the use of long-range classical communication, this is not part of our scheme; nonetheless, we still obtain a dramatic advantage.
For our setup, we assume that at $t=0$ we are in a state of fixed parity. The Hamiltonian interaction can be implemented using two unitaries in a brickwork structure [see \cref{fig:8}(a)]:
\begin{align}
    \hat U_{H,1} &= \exp \left( -iJ \sum_{s = A,B} \sum_{i = 1} \hat \sigma_{s,2i-1}^+ \hat \sigma_{s,2i}^- - \hc \right), \nonumber \\
    \hat U_{H,2} &= \exp \left(-iJ \sum_{s = A,B} \sum_{i = 1} \hat \sigma_{s,2i}^+ \hat \sigma_{s,2i + 1}^- - \hc \right), \label{eqn:circuitGatesHam} 
\end{align}
which are independent of the classical parity variable $P$. To implement the quantum jumps, we use two other unitaries as well as an auxilliary qubit $\hat \sigma_{\mathrm{aux}}$:
\begin{align}
    \hat U_{\mathrm{jump},1} &= \exp \left( -i \hat \sigma_{\mathrm{aux}}^+ \left[ u \hat \sigma_{A,1}^- - v P \hat \sigma_{B,1}^+  \right] - \hc \right), \nonumber \\
    \hat U_{\mathrm{jump},2} &= \exp \left( -i \hat \sigma_{\mathrm{aux}}^+ \left[ u  \hat \sigma_{B,1}^- - v P\hat \sigma_{A,1}^+  \right] - \hc \right). \label{eqn:circuitGatesJump}
\end{align}
We imagine initializing the auxiliary qubit in $|0 \rangle$ and then acting with $\hat U_{\mathrm{jump},1}$. Then we perform a projective measurement on the auxiliary qubit in the computational basis. If the site is still in $|0\rangle$ then no jump occurred and we do nothing. If the site is now in $|1\rangle$, then we acted one time with $\hat L_1$, and so we have done a single jump. We update the parity $P \to -P$, and then reset the auxiliary qubit. We repeat the same process with $\hat U_{\mathrm{jump},2}$. 
This type of measurement and reset operation is a well known method to simulate Lindblad dynamics on a unitary circuit (see e.g.~\cite{LLoyd2001,Shen2017,Han2021}). However, using it within this kind of adaptive setting (where the subsequent unitary evolution is updated based on the presence or absence of a jump) was not considered.  

The above gates define one layer of an adaptive circuit [see \cref{fig:8}(a)] which implements a time-discretized version of the fermionic master equation on the spins, and therefore can converge to the proper steady state without being constrained by time-entanglement trade-offs. By tracking parity and using it to adapt the dynamics, we are effectively trading a long range spatial interaction for a minimal kind of non-Markovian dynamics (using a single-bit classical memory). The entanglement dynamics of this circuit (averaged over measurement outcomes) is shown in \cref{fig:8}(b), where we always start from an all-down product state. The bipartite entanglement entropy saturates to a maximal value in a time that grows {\it linearly} with system size, even for $v^2 = 0.5$ when the fixed point of the circuit generates $n$ perfect bell pairs (a maximally entangled state).  It thus evades the general time-entanglement bound of Ref.~\cite{Pocklington2024}. The difference in dynamics is even more remarkable at the single trajectory level: we find that for any given stochastic trajectory, the entanglement entropy increases monotonically with time, see \cref{fig:8}(c). We can understand this mechanistically by thinking of the dissipation as injecting entanglement into the system: by flipping the phase of each injected Bell pair, it is impossible to inadvertently remove entanglement from the system, see \cite {Supplement} for more details. 

To emphasize the importance of the adaptive dynamics, we can compare against a circuit that is identical except that the relative phase $P$ is not updated, but instead held constant in time.  For any value of $v^2 \neq 0.5$, the resulting dynamics still converges to the same entangled state as the adaptive protocol, but on a much longer time scale, one that diverges as $v^2 \to 0.5$. This is shown in \cref{fig:8}(d). The striking difference in dynamics can also be seen at the trajectory level, where instead of growing monotonically the entanglement entropy rapidly fluctuates in both directions before finally reaching the fixed point on an exponentially longer time scale, see \cref{fig:8}(c).

Additionally, we find that this scheme is surprisingly robust to measurement errors in the auxiliary qubit, see \cite{Supplement} for details.

\textit{Time continuous adapative dynamics---}
%
Our circuit-based adaptive protocol can be directly generalized to a continuous-time, fully autonomous stabilization scheme by simply promoting the classical memory used above to a quantum degree of freedom.  Here, we will use the operator $\hat \sigma_P^z$ of a single auxiliary qubit to effectively track jumps and to update the phase of the dissipator accordingly.  We thus consider a modified Lindblad master equation for our double spin-chain system, where the jump operators in \cref{eqn:fermi_spin_2} are now:   
\begin{align}
\begin{array}{c}
    \hat L_1 = \hat \sigma_P^x (u \hat \sigma_{A,1}^- + v \hat \sigma_{B,1}^+ \hat \sigma_P^z), \\
    \hat L_2 = \hat \sigma_P^x (u \hat \sigma_{B,1}^- \hat \sigma_P^z + v \hat \sigma_{A,1}^+).
\end{array}
\label{eqn:time_cts_jumps}
\end{align}
Each jump now flips the state of the $P$ qubit, and its state also determines the phase of the dissipators as needed.  We find that the master equation generated by these dissipators and the Hamiltonian in \cref{eqn:fermi_spin_1} can rapidly stabilize an entangled state, completely evading the general time-entanglement bound of Ref.~\cite{Pocklington2024}, see \cref{fig:4}.  One appealing way to understand this is that the $P$ spin acts like a classical memory, making the effective dynamics of the spin chains non-Markovian (and hence making the general bound inapplicable). While we have focused on a specific setup, the construction here can be used to accelerate any scheme using dissipators of the form in \cref{eqn:locality_constraint}.  The result is dissipative stabilization of arbitrary pure many-body entangled states without any time-entanglement tradeoff \cite{Supplement}. Moreover, we show that this is possible using a minimal non-local resource of a single auxiliary qubit coupled to each system.

\textit{Accelerated spin squeezing---}
Our approach can also accelerate dissipative stabilization schemes very different from those in \cref{eqn:XXZs,eqn:local_jump}.  For example, consider dissipative preparation of a spin squeezed state of $N$ two level systems (collective spin operators $\hat S^{x,y,z} = \sum_{i = 1}^N \hat \sigma_{i}^{x,y,z}$). The master equation 
\begin{align}
    \partial_t \hat \rho &= \Gamma \D[\hat S^+ + \tanh(r) \hat S^-] \hat \rho \label{eqn:spin_sq_std}
\end{align}
conserves total angular momentum $S_{\rm tot}$, and in the maximum $S_{\rm tot}$ subspace (and for $N$ even) has a unique, pure spin squeezed steady state.  For large $r$, this state has a Wineland parameter that exhibits Heisenberg limited scaling: $\xi^2_R \equiv N\langle (\hat S^x)^2 \rangle/|\langle \hat S \rangle |^2 \propto 2/(N+2)$, see  \cite{Agarwal1989,DallaTorre2013,Groszkowski2022} also \footnote{We also note that for this particular state, a Ramsey measurement is in fact optimal, and so the Quantum Fisher Information $\mathcal{F}_Q$ can be simply related to the Wineland Parameter via $\mathcal{F}_Q = N/\xi_R^2$. See \cite{Supplement} for more details.}. This scheme also has a slow-down problem, and a time-squeezing tradeoff:  as $r$ increases, the dissipative gap (relaxation rate) is observed to vanish as $\sim e^{-4r}$. The reason for this slow down is different than the spin chains, in that it is not directly related to entanglement. Instead, it is a result of a general constraint on dissipative preparation of low fluctuation states. In fact, recalling the fidelity to the steady state $F(t)$, for a Lindbladian with a jump operator $\hat L$, one can show that \footnote{While this expression does not obviously reflect the gauge invariance $\hat L \to e^{i \phi} \hat L$, it can be shown that, because the steady state is pure and annihilated by the jump operator, the variance of the real part of $\hat L$ is invariant under multiplication by a phase. See \cite{Supplement} for more details.}
\begin{align}
    \partial_t F \leq \langle (\hat L + \hat L^\dagger)^2 \rangle - \langle \hat L + \hat L^\dagger  \rangle^2,
\end{align}
and so the time scale is bounded by the fluctuations in the real part of the jump operator. In this case, because $\hat L + \hat L^\dagger \propto \hat S^x$ is being squeezed, it causes the dynamics to slow down. See \cite{Supplement} for details. 

\begin{figure}[t]
    \centering
    \includegraphics[width = \linewidth]{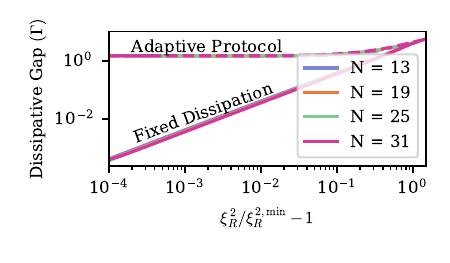}
    \caption{Dependence of the dissipative gap on the amount of squeezing. Each curve shows a fixed system size $N$, where the x-axis parameterizes how close the steady state Wineland squeezing parameter $\xi_R^2$ is compared to its minimal value $\xi_R^{2,\mathrm{min}}$, and is tuned using the squeezing parameter $r$. Solid lines correspond to the standard protocol with fixed dissipation [c.f.~\cref{eqn:spin_sq_std}], and dashed lines are the adaptive protocol where the phase flips after every jump.}
    \label{fig:10}
\end{figure}

We can use the above fermion-inspired methods to modify and accelerate the dynamics here in a manner than preserves the steady state squeezing. It is rather unintuitive that this works - it means that after every jump, the squeezing axis switches from $\hat S^x \leftrightarrow \hat S^y$. It seems like half the time, the jump operator is actually trying to stabilize the wrong steady state. However, as can be seen in \cref{fig:10}, this dramatically increases the relaxation rate. This is reminiscent of the quantum Mpemba effect, where starting farther away from the steady state allows you to reach it more quickly \cite{Mpemba1969,Carollo2021,Chatterjee2023}.

In order to actually implement this, one approach would be to use the ``discrete-time'' method. If dissipative squeezing is achieved via red- and blue-sideband drives on a cavity \cite{DallaTorre2013}, then one can keep track of the quantum jumps by doing single photon detection on the cavity output field. Then, one can update the jump operator $\D[\hat S^+ + \tanh(r) \hat S^-] \leftrightarrow \D[\hat S^+ - \tanh(r) \hat S^-]$ every time a photon is detected by toggling the phase of the drive, see \cite{Supplement} for details. Alternatively, the continuous-time approach would use the modified master equation
\begin{align}
    \partial_t \hat \rho &= \Gamma \D \left[ \hat \sigma^x_{\mathrm{aux}} \left( \hat S^+ + \hat \sigma^z_\mathrm{aux}\tanh(r) \hat S^- \right) \right] \hat \rho. \label{eqn:spin_sq_adapt}
\end{align}
Here, $\hat \sigma^z_\mathrm{aux}$ corresponds to an auxiliary qubit degree of freedom. Surprisingly, either method can significantly improve the time scale on which squeezing occurs: the dissipative gap remains $\mathcal{O}(\Gamma)$ even for arbitrarily large $r$ and maximal squeezing. 

\textit{Conclusion---}
We have demonstrated how an extremely minimal kind of adaptive dynamics can be used to dramatically speed up the stabilization of highly entangled many-body states.  This includes the generation of remote entanglement, as well as the stabilization of spin squeezed states.  Our protocols do not need long-range interactions, but nonetheless take direct inspiration from fermionic stabilization protocols, which are surprisingly immune from time-entanglement tradeoffs due to a subtle kind of in-built non-locality. In future work, it would be interesting to explore whether similar adaptive protocols could be used to speed up the preparation of other metrologically useful states.

This work was supported by the the Army Research Office under Grant No. W911NF-23-1-0077, the National Science Foundation QLCI HQAN (NSF Grant No. 2016136), the Air Force Office of Scientific Research MURI program under Grant No. FA9550- 19-1-0399, and the Simons Foundation through a Simons Investigator Award (Grant No. 669487).

\let\oldaddcontentsline\addcontentsline
\renewcommand{\addcontentsline}[3]{}

\bibliography{refAdaptiveDynamics}
\let\addcontentsline\oldaddcontentsline

\newpage 
\clearpage
\thispagestyle{empty}
\onecolumngrid
\begin{center}
\textbf{\large Supplemental Material for\\Accelerating Dissipative State Preparation with Adaptive Open Quantum Dynamics }
\end{center}

\begin{center}
Andrew Pocklington,${}^{1,2}$ and Aashish A. Clerk${}^2$\\
\emph{
${}^1$Department of Physics, University of Chicago, 5640 South Ellis Avenue, Chicago, Illinois 60637, USA  \\
${}^2$Pritzker School of Molecular Engineering, University of Chicago, Chicago, IL 60637, USA} \\
(Dated: \today)
\end{center}

\setcounter{equation}{0}
\setcounter{figure}{0}
\setcounter{table}{0}
\setcounter{page}{1}
\makeatletter
\renewcommand{\theequation}{S\arabic{equation}}
\renewcommand{\thefigure}{S\arabic{figure}}
\renewcommand{\bibnumfmt}[1]{[S#1]}
\renewcommand{\citenumfont}[1]{#1}


\tableofcontents

\section{Monotonicity of Entanglement in the Monitored Circuit}

We will now show that in the monitored circuit, by flipping the phase after every jump, the entanglement grows monotonically in time. This can be understood by looking at the interplay of the coherent and dissipative dynamics. To see this, it will be useful to rewrite the dynamics as
\begin{align}
    \hat H &= \sum_{i = 1}^{n-1} \hat h_i \\
    \hat h_i &= J \sum_{s = A,B}  \hat \sigma_{s,i}^+ \hat \sigma_{s,i + 1}^- + \hc \\
    \hat L_\pm &= \left( \hat \sigma_{A,1}^+ \pm \hat \sigma_{B,1}^+ \right)/\sqrt{2}.
\end{align}
When the parity is $P = +1$, then the dynamics are given by $\partial_t \hat \rho = -i[\hat H, \hat \rho] + \D[\hat L_+]\hat \rho + \D[\hat L_+^\dagger] \hat \rho$. When the parity is $P = -1$, the jump operators are exchanged $\hat L_+ \leftrightarrow \hat L_-$, while the Hamiltonian is unchanged. Note that this is identical to the dynamics presented in Eqs.~(6)~and~(7) in the main text (where we have taken maximal entanglement $u^2 = v^2 = 0.5$), after performing a unitary transformation inverting every site on the $B$ chain: $\hat \sigma_{B,i}^+ \leftrightarrow  (-1)^{i + 1} \hat \sigma_{B,i}^-$. The phase factor $(-1)^{i+1}$ makes it so both lattices have positive hopping terms. Following this transformation, the steady state becomes
\begin{align}
    \rss &= 2^{-n/2} \prod_{i = 1}^{n} \left( \hat \sigma_{A,i}^+ + (-1)^{i + 1} \hat \sigma_{B,i}^+ \right) |0 \rangle^{\otimes 2n}.
\end{align} 
Next, it will be useful to group the $A$ and $B$ lattices together into a single chain of 4-level systems. We define the following four states:
\begin{align}
\begin{array}{c}
|00\rangle_j  = |0\rangle_{A,j} \otimes |0 \rangle_{B,j}, \ \ \ \ \ |11\rangle_j  = |1\rangle_{A,j} \otimes |1 \rangle_{B,j} \\
\\
|T/S \rangle_j  = \frac{1}{\sqrt{2}} \left( |0\rangle_{A,j} \otimes |1 \rangle_{B,j} \pm  |1\rangle_{A,j} \otimes |0 \rangle_{B,j} \right),
\end{array}
\end{align}
In this basis, the (positive parity) steady state is simply $|T\rangle_1 |S\rangle_2 \dots |T/S\rangle_{n}$. We have the following rules for local interactions under the local Hamiltonian $\hat h_i$:
\begin{align}
\begin{array}{c}
|S\rangle_i |00\rangle_{i +1} \leftrightarrow |00\rangle_i |S\rangle_{i +1}, \ \ \ \ |T\rangle_i |00\rangle_{i +1} \leftrightarrow |00\rangle_i |T\rangle_{i +1} \\
\\
|S\rangle_i |11\rangle_{i +1} \leftrightarrow |11\rangle_i |S\rangle_{i +1}, \ \ \ \ |T\rangle_i |11\rangle_{i +1} \leftrightarrow |11\rangle_i |T\rangle_{i +1}.
\end{array}
\end{align}
Taking inspiration from Refs.~\cite{lingenfelter2023,Znidaric2013}, we can interpret $|00\rangle_1 |00\rangle_2 \dots |00 \rangle_n$ as a vacuum state with singlets $|S\rangle$ and triplets $|T\rangle$ hopping ballisticly under the Hamiltonian. Furthermore, the following are all zero energy eigenstates of the local Hamiltonian $
\hat h_i$:
\begin{align}
|00\rangle_i |00 \rangle_{i + 1},& \ \ \ \ |11\rangle_i | 11 \rangle_{i + 1}, \nonumber \\
|S\rangle_i | T \rangle_{i + 1},& \ \ \ \ |T\rangle_i | S \rangle_{i + 1}. \nonumber
\end{align}
Finally, $|S\rangle$ and $|T\rangle$ particles are able to scatter off of themselves under the local Hamiltonian $\hat h_i$:
\begin{align}
\begin{array}{c}
|S\rangle_i | S \rangle_{i + 1} \leftrightarrow -(|00\rangle_i | 11 \rangle_{i + 1} + |11\rangle_i | 00 \rangle_{i + 1}), \\
\\
|T\rangle_i | T \rangle_{i + 1} \leftrightarrow (|00\rangle_i | 11 \rangle_{i + 1} + |11\rangle_i | 00 \rangle_{i + 1}),
\end{array}
\end{align}
and so there is a second order process able to convert $|S\rangle_i | S \rangle_{i + 1}$ to $|T\rangle_i | T \rangle_{i + 1}$ and vice-versa.

\begin{figure}[t]
\centering
\includegraphics{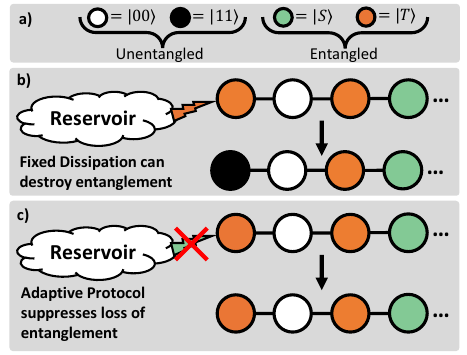}
\caption{ 
a) Legend depicting different states that can exist in the doubled chain structure. The white/black states ($|00\rangle, |11\rangle)$ are unentangled, while the colored Singlet/Triplet states carry maximal entanglement.
b) For the non-adaptive protocol with fixed jump operators, after inserting a triplet $|T\rangle$, it can return to the dissipative site, after which trying to insert a second $|T\rangle$ converts it to $|1\rangle$, removing entanglement.
c) Using the adaptive approach, because $|T\rangle,|S\rangle$ cannot cross, after inserting a $|T\rangle$, only a $|T\rangle$ can return to the dissipative site. However, inserting now an $|S\rangle$ on top of a $|T\rangle$ does nothing, and so the adaptive protocol supresses the entanglement loss. Hence, entanglement must grow monotonically in time in each stochastic unraveling.
}
\label{fig:mono}
\end{figure}

Now the last question to ask is what the dissipation can do. We have the next set of rules:
\begin{align}
\hat L_{+} |00 \rangle_1 &= \hat L_{+}^\dagger |11\rangle_1= |T \rangle_1, \\
\hat L_{-} |00 \rangle_1 &= \hat L_{-}^\dagger |11\rangle_1= |S \rangle_1, \\
\hat L_{+} |T \rangle_1 &= - \hat L_{-} |S \rangle_1 = |11 \rangle_1, \\
\hat L_{+}^\dagger |T \rangle_1 &= - \hat L_{-}^\dagger |S \rangle_1 = |00 \rangle_1,
\end{align}
with everything else annihilated. Suppose for convenience that the system begins in the vacuum of all $|00\rangle$, and we initialize to positive parity. The Hamiltonian does nothing until a quantum jump occurs, at which point a triplet $|T\rangle_1$ can be inserted in the boundary of the chain, which can then hop ballisticly under the Hamiltonian. Now we imagine two scenarios: first, suppose that we do not update the parity, and so the jump operator is unchanged. Thus, a second jump will again try to insert a second triplet $|T\rangle_1$. Now, if the first triplet has propagated away, this will be successful, and entanglement entropy will increase. However, if the original particle is either still at the first site, or has propagated and away and then returned, a quantum jump can accidentally \textit{destroy} the triplet as $\hat L_+ |T\rangle_1 = |1 1\rangle_1$ and $\hat L_+^\dagger |T\rangle_1 = |00\rangle_1$, both of which are unentangled, see \cref{fig:mono}. Note that the total amount of entanglement between the two chains can only change at the boundary, as the Hamiltonian does not contain any interactions between the $A$ and $B$ subsystems. 

In contrast, consider the dynamics when we do update the jump operators after every jump (i.e.~update the $P$ variable).  After the first jump (which injected a $|T \rangle$, the jump operator in the master equation switches to being $\hat L_-$. Now, a quantum jump tries inject a singlet into the chain. Again, if the original excitation has propagated away leaving vacuum at the first site, this will increase the entanglement entropy. On the other hand, if the original triplet has either not yet left or has come back, we can observe that $\hat L_-^{(\dagger)} |T\rangle_1 = 0$, and so a quantum jump \textit{does nothing}. Because an alternating pattern of singlets and triplets cannot cross places or scatter off of each other, if the last type inserted was a singlet, only a singlet can come back to the first site, and so by updating the parity after every jump it is impossible to remove entanglement from the system. This process is shown in \cref{fig:mono}. An identical argument holds if one instead considers the $|11\rangle_1 \dots |11\rangle_n$ state to be the vacuum, where the rolls of $\hat L_\pm$ and $\hat L_\pm^\dagger$ are simply reversed.

Thus, we have shown monotonicity of the entanglement entropy - it only grows before saturating to a maximal value. This is shown at the individual trajectory level, where we compare the scheme where we keep track of jumps (variable phase) and where we do not (fixed phase). The variable phase shows monotonic entanglement growth that converges orders of magnitude more quickly, while the fixed phase dynamics take a very long time to reach the steady state and shows oscillating entanglement. See Fig.~2(c) in the main text.

\section{Circumventing Slowdown in Random, Many-Body Lindbladians}

\begin{figure}[t]
    \centering
    \includegraphics{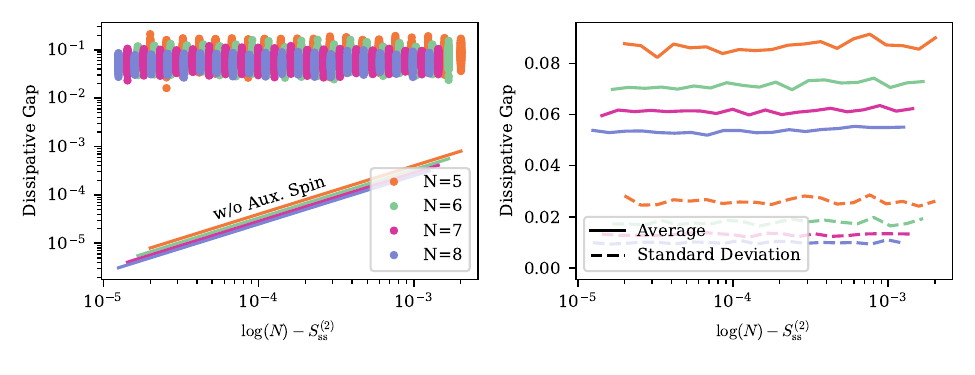}
    \caption{Left. Random Lindbladians with fixed Renyi-2 steady state entanglement constructed using an auxiliary spin [c.f.~\cref{eqn:aux_spin_jump,eqn:aux_spin_L}]. 250 runs are shown for each value of the entanglement $S^{(2)}_\mathrm{ss}$ and system size $N$. Solid lines are analytical results for random Lindbladians without an auxiliary spin, see \cite{Pocklington2024}. Right. Average and standard deviation of the runs.}
    \label{fig:9}
\end{figure}

The construction given in Eq.~(12) in the main text can be straightforwardly generalized to any many-body system. In particular, recall the construction presented in Ref.~\cite{Pocklington2024}: given a bipartite Hilbert space $\mathcal{H} = \mathcal{H}_A \otimes \mathcal{H}_B$ of equal Hilbert space dimension $\dim \mathcal{H}_A = \dim \mathcal{H}_B = N$, then a quantum state is uniquely specified by a set of real, positive Schmidt coefficients $\psi_i$ and basis vectors $|i\rangle$:
\begin{align}
    |\psi \rangle &= \sum_{i= 1}^N \psi_i |i\rangle \otimes |i \rangle.
\end{align}
Now, given any operator $\hat A_\mu \in \mathrm{End}\left( \mathcal{H}_A \right)$, we can construct a jump operator $\hat L_\mu$ such that $\hat L_\mu |\psi \rangle = 0$ by
\begin{align}
    \hat L_\mu &= \hat A_\mu \otimes \1 - \1 \otimes \hat \Psi \hat A^T_\mu \hat \Psi^{-1}.
\end{align}
Here, $\hat \Psi = \sum_i \psi_i |i \rangle \langle i | = \sqrt{\tr_A |\psi \rangle \langle \psi |}$. The transposition is short hand for $\hat A^T = \hat \Theta \hat A^\dagger \hat \Theta^{-1}$ where $\hat \Theta$ is an anti-unitary time reversal operation (i.e., complex conjugation in the Schmidt basis).

Now, one can choose a set of $N$ random Schmidt coefficients $\psi_i$ (subject to normalization) to define a steady state $|\psi \rangle$, as well as $N^2$ random matrix elements $(A_\mu)_{ij}$ to define
\begin{align}
    \hat A_\mu = \sum_{i,j=1}^N (A_\mu)_{ij} |i \rangle \langle j |.
\end{align}
The construction then allows one to construct from this a Lindbladian
\begin{align}
    \L &= \sum_{\mu = 1}^m \D[\hat L_\mu]
\end{align}
where $\hat L_\mu|\psi \rangle= 0 $ by construction. As long as $m > 1$, this state is also generically unique; however, it was shown in \cite{Pocklington2024} that the dissipative gap closes as the steady state approaches maximal entanglement $S^{(2)}_\mathrm{ss} \to \log(N)$.

In order to still reach the state $|\psi \rangle$ in finite time, we can now use the construction in Eq.~(12) in the main text and augment the Hilbert space to $\mathcal{H} = \mathcal{H}_\mathrm{aux} \otimes \mathcal{H}_A \otimes \mathcal{H}_B$ where $\dim \mathcal{H}_\mathrm{aux} = 2$, a single auxiliary spin degree of freedom. Now consider the new state and jump operators
\begin{align}
    |\psi_2 \rangle &= |0 \rangle \otimes |\psi \rangle, \\
    \hat L_{\mu,2} &= \hat \sigma_x \otimes \hat A_\mu \otimes \1 + i \hat \sigma_y \otimes \1 \otimes \hat \Psi \hat A^T_\mu \hat \Psi^{-1}. \label{eqn:aux_spin_jump}
\end{align}
it is simple to check that, by construction, $\hat L_{\mu,2} |\psi_2 \rangle = 0$, and it is once again the unique fixed point of the Lindbladian
\begin{align}
     \L_2 &= \sum_{\mu = 1}^m \D[\hat L_{\mu,2}], \label{eqn:aux_spin_L}
\end{align}
but $\L_2$ does not experience slow-down as $|\psi \rangle$ approaches maximal entanglement. Indeed, as shown in \cref{fig:9}, the average dissipative gap with the auxiliary spin is independent of steady state entanglement.

\section{Spin Squeezing}

\subsection{Squeezing Induced Slowdown}

Here we discuss how the adaptive protocol can be used for spin squeezing. First, it is important to understand exactly why the dynamics become slow in the large squeezing limit. This is not related to the entanglement (as it happens within a single ensemble) but instead can be related directly to the amount the variance is being squeezed. 

Let's consider an extremely general scenario, such that there are dissipative dynamics 
\begin{align}
    \partial_t \hat \rho &= \D[\hat L] \hat \rho,
\end{align}
with a pure steady state $\rss = |\psi \rangle \langle \psi |$. 
Now, let's rewrite $\hat L$ as its hermitian and anti-hermitian parts:

\begin{align}
\hat L &= \hat O_1 + i \hat O_2,
\end{align}
where $\hat O_i = \hat O_i^\dagger$. Now, observe that the steady state condition tells us that $\hat L |\psi \rangle = 0$, so
\begin{align}
\hat O_1 |\psi \rangle &= -i \hat O_2 |\psi \rangle. \label{eqn:O1O2}
\end{align}
However, this also implies that $\var(\hat O_1) = \var( \hat O_2)$ where 
\begin{align}
\var(\hat O) \equiv \langle \psi | \hat O^2 | \psi \rangle - \langle \psi | \hat O | \psi \rangle^2.
\end{align}
Now, we can use Heisenberg uncertainty to observe that 
\begin{align}
\var(\hat O_1)^2 &= \var(\hat O_1) \var(\hat O_2) \geq \frac{1}{4} |\langle [\hat O_1, \hat O_2] \rangle |^2 = \frac{1}{16} | \langle [\hat  L, \hat L^\dagger ] \rangle |^2.
\end{align}
However, we recall from \cite{Pocklington2024} that if we define $F(t) = \tr (\hat \rho_t \rss)$ the fidelity to the steady state at time $t$, then 
\begin{align}
|\partial_t F|^2 & \leq |\langle [\hat L, \hat L^\dagger] \rangle|^2. \\
\implies \partial_t F & \leq 4\var( \hat O_1 ).
\end{align}
At this point, it does not seem that this respects the gauge invariance of the jump operator $\hat L \to e^{i \phi} \hat L$ as such a transformation changes the definition of $\hat O_1 \to \cos(\phi) \hat O_1 - \sin(\phi) \hat O_2$. However, it is simple to show that if we consider
\begin{align}
    \left\langle \left( \cos(\phi) \hat O_1 - \sin(\phi) \hat O_2 \right)^2 \right\rangle &= \cos^2 (\phi) \langle   \hat O_1^2 \rangle + \sin^2 (\phi) \langle   \hat O_2^2 \rangle - \sin(\phi) \cos(\phi) \langle \{ \hat O_1, \hat O_2 \} \rangle \nonumber \\
    &= \langle \hat O_1^2 \rangle,
\end{align}
where we use \cref{eqn:O1O2} to see that $\langle \{ \hat O_1, \hat O_2 \} \rangle = 0$. \cref{eqn:O1O2} also implies that $\langle \hat O_1 \rangle = -i\langle \hat O_2 \rangle$ and so they must both be zero (as one is purely real and the other purely imaginary). Hence, this proves the variance of the real part of the jump operator is in fact gauge invariant.

Now, if we apply this to the case of dissipative spin-squeezing, we can observe that
\begin{align}
    \hat L &= \hat S^+ + \tanh(r) \hat S^- = (1 + \tanh(r)) \hat S^x + i(1 - \tanh(r)) \hat S^y,
\end{align}
which implies that, in the large $r$ limit $e^{2r} \gtrsim N$ that
\begin{align}
    \partial_t F & \leq 4(1 - \tanh(r))^2 \var(\hat S^y) \sim 2S(S + 1) e^{-4r},
\end{align}
and hence the system slows down at large levels of squeezing. Note that because they are equal, one could also have shown the same slowdown by considering the variance $(1 + \tanh(r))^2 \var (\hat S^x)$. This also allows us to understand why the auxiliary qubit can speed up the dynamics. If we write the adaptive master equation
\begin{align}
    \partial_t \hat \rho &= \D[\hat{\tilde L}] \hat \rho, \\
    \hat{\tilde L} &= \hat \sigma^x \hat S^+ + i \tanh(r) \hat \sigma^y \hat S^- = \left( \hat \sigma^x \hat S^x + \tanh(r) \hat \sigma^y \hat S^y \right) + i \left( \hat \sigma^x \hat S^y + \tanh(r) \hat \sigma^y \hat S^x \right).
\end{align}
The steady state solution is unchanged, but the variance now becomes
\begin{align}
    &\var \left( \hat \sigma^x \hat S^x + \tanh(r) \hat \sigma^y \hat S^y \right) = \left\langle (\hat S^x)^2 + \tanh^2(r) (\hat S^y)^2 + 2 \tanh(r) \hat \sigma^z \hat S^z \right\rangle
\end{align}
In the limit $r \to \infty$, the steady state is an $\hat S^x$ eigenstate with eigenvalue $0$. Hence, we have that $\langle (S^x)^2 \rangle \to 0$ (as it is being squeezed). However, because $\hat S \cdot \hat S = S(S + 1)$ where $S$ is the total angular momentum, then by rotational invariance we must conclude that $\langle (\hat S^y)^2 \rangle = \langle (\hat S^z)^2 \rangle = S(S + 1)/2$. Finally, noting that an $\hat S^x$ eigenstate similarly implies $\langle \hat S^z \rangle = 0$, we can conclude
\begin{align}
    \var \left( \hat \sigma^x \hat S^x + \tanh(r) \hat \sigma^y \hat S^y \right)\bigg|_{r \to \infty} &= \frac{S(S + 1)}{2},
\end{align}
Thus, the system no longer has to suffer from slow down, as was seen in the main text. 

\begin{figure}[t!]
    \centering
    \includegraphics{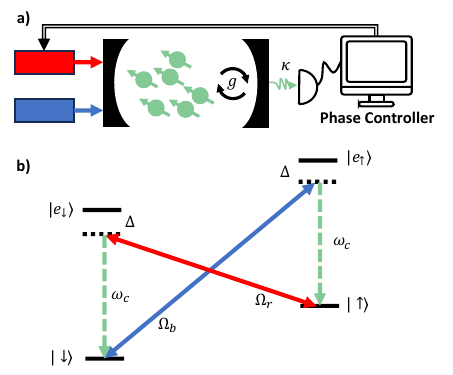}
    \caption{a) Adaptive spin squeezing protocol. Red and Blue sideband drives on a cavity QED setup generate dissipative squeezing. The output field of the cavity is measured by a single photon detector, which is fed forward to control the relative phases of the drives. b) Level diagram of four level atoms. Sideband drives excite from the ground to excited state manifold, which can then decay back to the ground state via emission into the cavity field at frequency $\omega_c$.}
    \label{fig:adaptiveSq}
\end{figure}

\subsection{Quantum Fisher Information}
Throughout, we have discussed the metrological gain of the spin squeezed states in terms of their Wineland squeezing parameter, which gives the degree to which the spin squeezed state will outperform the standard quantum limit when doing a simple Ramsey measurement. Now, another manner to define the metrological use of a quantum state is via the quantum Fisher information (QFI), which gives the sensitivity to a sensing parameter optimizing over all measurement schemes.

As it turns out, for the dissipative squeezed states we have looked at, these are in one-to-one mapping, because the Ramsey measurement turns out to be optimal. There are a few ways in which one can see this. Perhaps the most straightforward is that we can directly calculate the symmetric logarithmic derivative (SLD). Let us define the squeezed state $\hat \rho = |\psi\rangle \langle \psi |$ as before as the kernel of the jump operator
\begin{align}
    (\hat S^+ + \tanh(r) \hat S^- ) |\psi \rangle &= 0 \\
    \implies \hat S^x |\psi \rangle &= i e^{4 r} \hat S^y |\psi \rangle
\end{align}
If we are using this to sense a parameter $\theta$ encoded as a Hamiltonian $\hat H = \theta \hat S^y$, then the SLD $\hat L_\theta$ is defined by
\begin{align}
    i[\hat \rho, \hat S^y] &= \frac{1}{2} \{ \hat L_\theta, \hat \rho \} \\
    \implies \hat L_\theta &= -2 e^{-4 r} \hat S^x
\end{align}
The QFI $\mathcal{F}_Q$ is given by the expectation value of the SLD squared, which is just
\begin{align}
    \mathcal{F}_Q &= 4 e^{-8 r} \langle (\hat S^x)^2\rangle = 4 \langle (\hat S^y)^2\rangle
\end{align}
Now, on the other hand, we can exactly calculate Wineland parameter using the Heisenberg uncertainty relation. The most general form of the uncertainty relation is given by
\begin{align}
    \sigma_A^2 \sigma_B^2 \geq \frac{1}{4} \left| \langle \{ \hat A, \hat B\} \rangle - \langle \hat A \rangle \langle \hat B \rangle \right|^2 + \frac{1}{4} \left| \langle [ \hat A, \hat B]  \rangle \right|^2
\end{align}
The inequality is saturated when the states $|\phi_A \rangle = (\hat A - \langle \hat A \rangle )|\psi \rangle$ and $|\phi_B \rangle = (\hat B - \langle \hat B \rangle )|\psi \rangle$ are parallel. Let's consider $\hat A = \hat S^x$ and $\hat B = \hat S^y$. Now, as we know the expectation of both these operators is 0, and therefore $|\phi_{S^x} \rangle = i e^{4r} |\phi_{S^y} \rangle$, and so the inequality is saturated. Further, we can observe that 
\begin{align}
    \langle \hat S^x \hat S^y \rangle &= -i e^{-4r} \langle (\hat S^x)^2 \rangle = -\langle \hat S^y \hat S^x \rangle \\
    \implies \{\langle \hat S^x, \hat S^y \} \rangle &= 0 \\
    \implies \sigma_{S^x}^2 \sigma_{S^y}^2 &= \frac{1}{4} \left| \langle [ \hat S^x, \hat S^y]  \rangle \right|^2 = \frac{1}{4} \left| \langle \hat S^z  \rangle \right|^2 \\
    \implies \mathcal{F}_Q &= 4 \sigma_{S^y}^2 = \frac{\left| \langle \hat S^z  \rangle \right|^2}{\sigma_{S^x}^2} = \frac{N}{\xi_R^2}
\end{align}
and so the QFI is simply proportional to the squeezing parameter.

\subsection{Experimental Implementation}

While this master equation seems to be difficult to construct in the lab, one could instead consider doing the adaptive technique. To engineer dissipative spin-squeezing, we consider the protocol in Ref.~\cite{DallaTorre2013}: we take a cavity with frequency $\omega_c$ and loss rate $\kappa$. Inside the cavity are a collection of $N$ 4-level atoms. The spin degree of freedom will be encoded in the ground state manifold $|\uparrow \rangle, | \downarrow \rangle$, and there are also two excited states $|e_\uparrow \rangle, |e_\downarrow \rangle$. We take an atom-cavity coupling $g$, along with blue- and red- sideband drives with strength $\Omega_b, \Omega_r$ at a detuning $\Delta$ from the $|\downarrow \rangle \leftrightarrow |e_\uparrow \rangle$ and $|\uparrow \rangle \leftrightarrow |e_\downarrow \rangle$ transitions, respectively. Assuming $\Omega_{b,r}/\Delta \ll 1$, we can trace out the excited states, and working in the proper rotating frame gives the effective Hamiltonian \cite{DallaTorre2013}:
\begin{align}
    \hat H &= \tilde{\Omega}_b \left( \hat a^\dagger \hat S^+ + \hat a \hat S^- \right) + \tilde{\Omega}_r \left( \hat a^\dagger \hat S^- + \hat a \hat S^+ \right),
\end{align}
where $\tilde{\Omega}_{b,r} = \Omega_{b,r}g/\Delta$ are effective Rabi rates \cite{DallaTorre2013}. If the cavity additionally is extremely lossy with a loss rate $\kappa \gg \tilde{\Omega}_b, \tilde{\Omega}_r$, then we arrive at the dissipative dynamics for just the spins as
\begin{align}
    \partial_t \hat \rho &= \frac{2}{\kappa}\D\left[ \tilde{\Omega}_b  \hat S^+ + \tilde{\Omega}_r \hat S^- \right] \hat \rho.
\end{align}
This gives us the dissipative dynamics necessary for spin squeezing. Finally, notice that we can keep track of the number of jumps that have occurred by doing single photon detection on the cavity output field. Every time we detect a photon, a jump occurred, and we can change the relative phase of the red and blue sideband drives to change the relative sign in the jump operator. This could be done using, e.g., electro-optic modulation of the red sideband, so that $\Omega_r \to -\Omega_r$. This process is depicted in \cref{fig:adaptiveSq}.

\section{Scalability}
In this section, we will investigate the scalability of both the proposed adaptive circuit protocol, as well as the adaptive spin-squeezing protocol. Specifically, we will be interested in understanding how the dissipative preparation time scales with the system size.

\begin{figure}[t]
    \centering
    \includegraphics{ 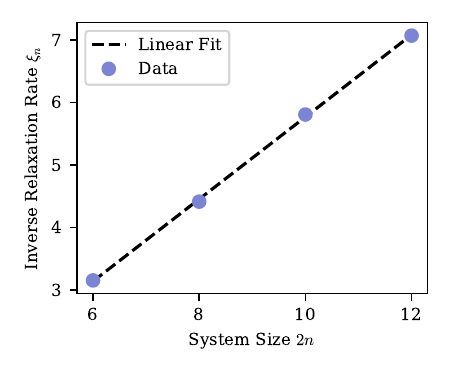}
    \includegraphics{ 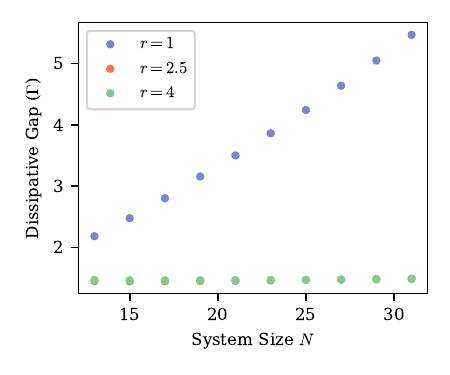}
    \caption{On the left, we show the average preparation time for the dissipative circuit [see \cref{eqn:invRelRate}] as a function of the system size $2n$, where the data points are obtained by fitting averaged bipartite entanglement to exponential decay curves. The linear fit is given by roughly $\xi_n \sim 0.66 \times  (2n)$. On the right, we numerically diagonalize the Liouvillian of the spin squeezing master equation and extract the dissipative gap, or smallest non-zero eigenvalue, which sets the relaxation time. }
    \label{fig:scalability}
\end{figure}

For the circuit, we can observe that the steady state has long-range entanglement between the most distant qubits in the $A$ and $B$ subsystems; however, the only location in which entanglement can be created between the two sub-systems is at the auxilliary lattice site. Hence, it is clear that in order to reach the fixed point, there must be at least linear-depth scaling, i.e. the number of cycles for the system to relax must scale at least as $n$, where $n$ is the number of qubits in each sub-system. Numerically, we find that this is accurate, as can be seen in \cref{fig:scalability}, where it is shown that in numerically accessible system sizes, the bipartite entanglement scales as
\begin{align}
    S_\mathrm{vN} &\sim n \left( 1 - e^{-d/\xi_n} \right) \label{eqn:invRelRate} \\
    \xi_n &\propto n
\end{align}
where $d$ is the number of circuit cycles, and $\xi_n$ the inverse relaxation rate. This means the actual system relaxes with an optimal scaling given a 1D connectivity.

For the spin-squeezed system, we expect two different behaviors of the relaxation rate, depending on the amount of squeezing. If the squeezing is small $e^{2r} \ll N$, then there is no slow-down to alleviate, and so the system should experience super-radiant decay, with a dissipative gap scaling linearly with $N$, the number of qubits. On the other hand, if $e^{2r} \gtrsim N$, then the super-radiant decay has been completely eliminated by the squeezing induced slow-down, and so the adaptivity is important to speed up the dynamics. However, the adaptivity does not induce super-radiance, and so the dissipative gap should be independent of both system size and squeezing. Both of these scalings are shown numerically in \cref{fig:scalability}. Hence, the protocol is extremely scalable, as the preparation time for highly squeezed states is independent of the number of particles.

\section{Measurement Error}

In this section, we investigate the impact of measurement errors on the adaptive protocols. In both the circuit as well as the spin-squeezing, the ability to do strong, projective measurements and then feed-back on them is crucial. As it turns out, in both scenarios, measurement errors essentially become leakage out of the correct parity sector, where here parity is a joint parity of the system plus classical memory.

\begin{figure}
    \centering
    \includegraphics{ 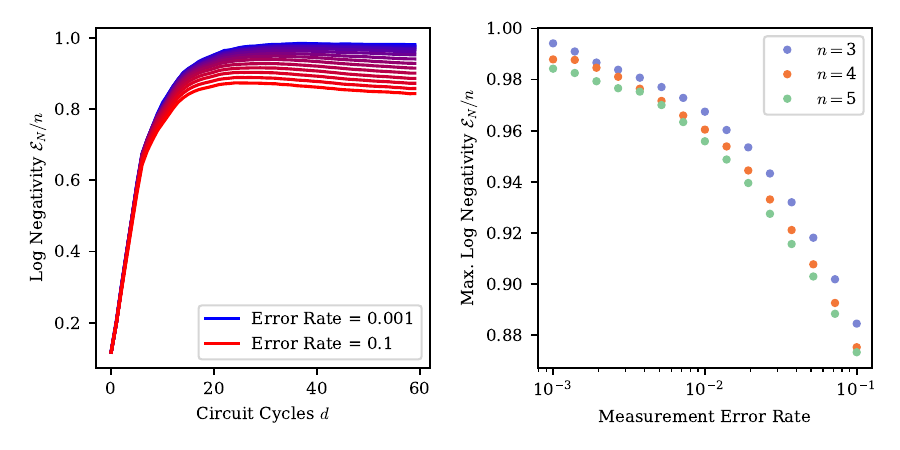}
    \caption{On the left, we plot the normalized Logarithmic Negativity per pair of qubits as a function of the number of circuit cycles $d$ for error rates between $0.1$ and $0.001$, and $n = 5$ qubit pairs. On the right is the maximal transient log negativity as a function of error rate for 3,4, and 5 pairs of qubits. Each point is averaged over 500 circuit trajectories.}
    \label{fig:circMeasErr}
\end{figure}

Let's begin by looking at the discrete circuit. The unitary operators $\hat U_{H,1}, \hat U_{H,2}$ conserve the total parity of the system. On the other hand, the unitary operators $\hat U_{\mathrm{jump},1}, \hat U_{\mathrm{jump},2}$ can change the parity of the system after performing a projective measurement on the auxiliary spin, which is stored in the classical memory. However, if there is a measurement error (i.e., we perform a projective measurement on the auxiliary site into $|1\rangle$ but record $0$ or vice-versa), then we have accidentally switched parity subspaces. By sampling over many circuit realizations, we can define the average, impure state as a statistical mixture
\begin{align}
    \hat \rho &= \lim_{m \to \infty} \frac{1}{m} \sum_{i = 1}^m |\psi^{(i)} \rangle \langle \psi^{(i)} |
\end{align}
where $m$ is the total number of circuit realizations, and the state $|\psi^{(i)}\rangle$ is the pure state of a single circuit realization. After many cycles, this can be written as
\begin{align}
    \hat \rho \to |\psi_\mathrm{ss} \rangle \langle \psi_\mathrm{ss} | + \hat \rho_\mathrm{err}
\end{align}
where $|\psi_\mathrm{ss} \rangle$ is the ideal steady state in the limit of perfect measurements, and $\hat \rho_\mathrm{err}$ is the population that has diffused into the incorrect subspace due to measurement errors. Now, because there is random flipping between the two at a rate given by the measurement error, in the limit of many cycles, the state will be fully degraded. However, if the rate at which measurement errors is much smaller than the number of cycles required to approach the ideal fixed point, there is a very long period during which the system nearly prepares the perfect state, before it is lost in the long time limit. Therefore, we can ask what the maximal amount of entanglement we prepare is after optimizing over circuit depth. We will take the logarithmic negativity as our mixed state entanglement monotone:
\begin{align}
    \mathcal{E}_N &= \log_2 ||\hat \rho^{T_A} ||_1
\end{align}
where $\hat \rho^{T_A}$ is the partial transpose of the $A$ system and $|| \hat O ||_1 = \tr \sqrt{\hat O^\dagger \hat O}$ is the one-norm. In \cref{fig:circMeasErr}, we can see that for a given measurement error rate, the log negativity peaks at nearly the maximal value, before slowly rolling off as measurement errors degrade the state. Moreover, we can see that even for a relatively significant error rate of 10\%, we can still achieve nearly 90\% of the optimal entanglement of $\mathcal{E}_N/n = 1$.

\begin{figure}
    \centering
    \includegraphics{ 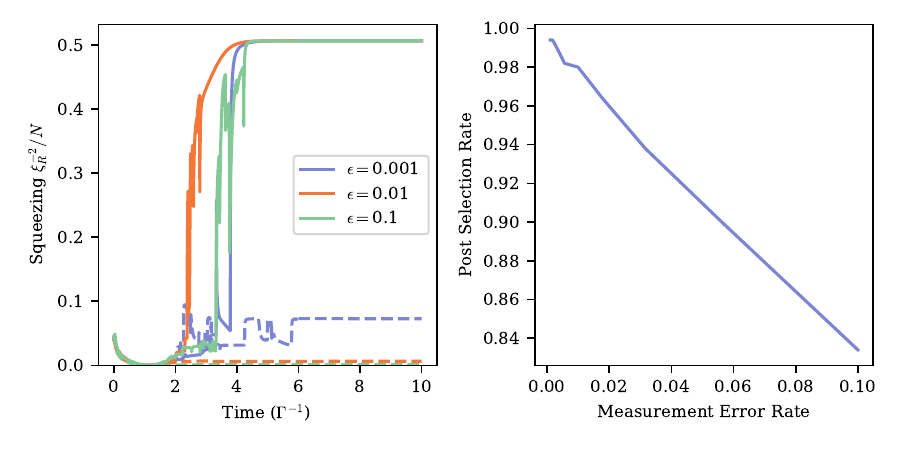}
    \caption{Left. Average Wineland squeezing parameter for measurement error rates $\epsilon = 0.1,0.01,0.001$ either with (solid) or without (dashed) post-selection. We have taken $N = 24$ spin ensemble, along with a moderate squeezing $r = 2$. On the right, we plot the fraction of trajectories that survive post-selection as a function of the measurement error rate. Plots averaged over 500 individual trajectories. }
    \label{fig:spinMeasErr}
\end{figure}

The dynamics for the spin squeezing example are very similar. Every time there is a quantum jump, the parity of the state flips, which is recorded in the classical register. If there is a measurement error, we will accidentally flip the global parity. The steady state in the opposite parity subspace is also spin squeezed; however, it squeezes an orthogonal quadrature of the spin ensemble. This is because every time we flip the sign of the classical register, we squeeze the wrong axis - thus, if we have an odd number of measurement errors, the long time state will be spin squeezed in the wrong axis. As it turns out, this is particularly troublesome for metrology, which can be seen from looking as the Wineland parameter. Let's consider two states $|\psi_X \rangle, |\psi_Y \rangle$ which are dissipatively spin squeezed along the $X$ and $Y$ directions, respectively. Then, if we let $\alpha$ be a parameter characterized by the measurement error rate, we can write the long time steady state as a statistical mixture
\begin{align}
    \hat \rho &= (1- \alpha) |\psi_X \rangle \langle \psi_X | + \alpha |\psi_Y \rangle \langle \psi_Y |
\end{align}
Now, if we are using $\hat \rho$ to do a Ramsey measurement, then we can characterize the sensitivity with the Wineland squeezing parameter
\begin{align}
    \xi_R^2 &= N \frac{\mathrm{var}(\hat S_x^2)}{|\langle \hat S \rangle|^2} 
\end{align}
Since both $|\psi_X \rangle$ and $|\psi_Y \rangle$ have the same expectation value of $\langle \hat S \rangle$, we can characterize the loss in sensitivity due to measurement error as
\begin{align}
    \frac{\xi_R^2(\alpha)}{\xi_R^2(\alpha = 0)} & = 1 - \alpha + \alpha \frac{\langle \psi_Y | \hat S_x^2 | \psi_Y \rangle}{\langle \psi_X | \hat S_x^2 | \psi_X \rangle} = 1 - \alpha + \alpha \frac{\langle \psi_X | \hat S_y^2 | \psi_X \rangle}{\langle \psi_X | \hat S_x^2 | \psi_X \rangle} \nonumber \\
    &= 1 - \alpha + \alpha \frac{(1 + \tanh r)^2}{(1 - \tanh r)^2} = 1 + (e^{4 r} - 1) \alpha
\end{align}
and so it appears that $\alpha$ must be exponentially small in order to have a state even be below the Standard Quantum Limit, let alone approach the Heisenberg limit. However, the situation is not in fact this dire. We can see that it is in fact only a few individual trajectories when we squeeze in the wrong direction that actually destroy the metrological gain of the entire ensemble; moreover, we always will know when they occur because we will always finish the protocol having measured an odd number of jumps. If we post-select on having performed an even number of quantum jumps at the end of the protocol, we will be able to keep the vast majority of the trajectories, and still have the ideal metrological sensitivity. This is shown in \cref{fig:spinMeasErr}. In the first plot on the left we show averaged dynamics for varying measurement error rates with either no post-selection (dashed line) or post selection (solid lines). Without post-selection, even for vanishing error rates of $0.01$, there is no squeezing. However, with post-selection, we immediately recover Heisenberg scaling for all error rates. In the second figure, we see that the post-selection rate is incredibly favorable, even for a measurement that is wrong 10\% of the time, we can keep more than 80\% of the individual runs, and so there is extremely little overhead.

\end{document}